\documentclass[cits]{PoS}
\usepackage{amsmath}
\usepackage{amssymb}
\usepackage{bbm}
\usepackage{caption}

\def \Nt {{N_{\tau}}}
\def \Nc {{N_{c}}}
\def \Nf {{N_{f}}}

\def \hmu {{\hat{\mu}}}

\def \bchi {{\bar{\chi}}}
\def \bpsi {{\bar{\psi}}}

\newcommand{\expval}[1]{\left\langle #1 \right\rangle}

\newcommand{\beqn} {\begin{equation}}
\newcommand{\eqn} {\end{equation}}

\def \beq{\begin{equation}}
\def \eeq{\end{equation}}
\def \bea{\begin{eqnarray}}
\def \eea{\end{eqnarray}}

\def \Tr {{\rm Tr}}
\def \tr {{\rm tr}}

\def \bet0{\beta_0}
\def \bet1{\beta_1}
\def \simgt{\,\rlap{\lower 7.5 pt\hbox{$\mathchar \sim$}}\raise 3 pt \hbox{$>$}\,}
\def \simlt{\,\rlap{\lower 7.5 pt\hbox{$\mathchar \sim$}}\raise 3 pt \hbox{$<$}\,}

\def\lsim{\raise0.3ex\hbox{$<$\kern-0.75em\raise-1.1ex\hbox{$\sim$}}}
\def\gsim{\raise0.3ex\hbox{$>$\kern-0.75em\raise-1.1ex\hbox{$\sim$}}}

\def \Id{\mathbbm{1}}

\newcommand{\Ord}[1]{\mathcal{O}\left( #1 \right)}

\title{Combinatorics of Lattice QCD at Strong Coupling}

\ShortTitle{Combinatorics of LQCD at Strong Coupling}

\author{\speaker{Wolfgang Unger}\\
Institut f\"ur Theoretische Physik, Goethe-Universit\"at Frankfurt,\\
60438 Frankfurt am Main, Germany\\
        E-mail: \email{unger@th.physik.uni-frankfurt.de}}

\abstract{
Thermodynamics in the strong coupling limit of lattice QCD has features which may be similar to those of continuum QCD, such as a chiral critical end point and a nuclear liquid gas transition. 
Here I compare the combinatorics of staggered and Wilson fermions in the strong coupling limit for arbitrary number of colors and flavors.
The partition functions can be considered as an expansions in hadronic spatial hoppings from the static limit, where both discretizations can be expressed via formulae with coefficients of distinct combinatorial interpretation. 
The corresponding multiplicites of hadronic states are evaluated using generalizations of Catalan numbers and Lucas polynomials.
I outline how quantum Monte Carlo simulations can be carried out in general, and summarize recent results on the gauge corrections to the strong coupling limit.
}

\FullConference{The 32nd International Symposium on Lattice Field Theory\\
                 23-28 June, 2014\\
                 Columbia University New York, NY}

\begin{document}

\section{Motivation}

The QCD phase diagram is conjectured to have a rich phase structure, but only little is known from lattice QCD due to the sign problem.
The available Monte Carlo methods are all limited to $\mu/T \lesssim 1$. Lattice QCD in the strong coupling limit,
$\beta=\frac{2\Nc}{g^2} \rightarrow 0$, and in a dual representation is a model where the sign problem is mild
enough to study the full $\mu$-$T$ phase diagram. 
This virtue crucially depends on the order of integration. The following three orders are common:
(1) Integrating out fermions first. This results in the fermion determinant $\det M[U]$. The Monte Carlo simulation is over gauge fields, $\beta$ can be varied continuously.
However, there is the severe sign problem at finite $\mu$, and it is expensive to approach the chiral limit.
(2) Integrating out spatial gauge links first then the fermions. The remaining temporal gauge links are mapped on Polyakov loops to obtain a 3-dim. heavy quark effective theory \cite{DePietri2007,Fromm2013}.
This is applicable to Wilson fermions, where backtracking of fermion world lines is prohibited. 
The fermion determinant is factorized into a kinetic and a static part. Corrections to the static limit are treated analytically (expansion in hopping parameter and gauge action up to some order $\mathcal{O}(\kappa^n u(\beta)^m))$.
(3) Integrating out all gauge links first, then the fermions. For staggered fermions, this leads to the Monomer-Dimer-System \cite{Rossi1984}, which has a mild sign problem, and the chiral limit is cheap.
There is no fermion determinant and it can be studied e.g.~via Worm algorithms \cite{Forcrand2010}.
For Wilson fermions, results only consist for the Schwinger model so far \cite{Salmhofer1991,Scharnhorst1996}.
Moreover, incorporating the gauge action requires additional gauge integrals and introduces plaquette occupation numbers \cite{Unger2014}.

I will focus here on strategy (3) and explain its combinatorial interpretation. Lattice QCD at strong coupling (SC-LQCD) shares important features with continuum QCD: 
it is ``confining'' in the sense that only color singlet d.o.f.~survive gauge integration, the mesons and baryons. These are point-like objects in the strong coupling limit, but become extended objects 
away from the strong coupling limit and mix with gluons. SC-LQCD also has a (nuclear) liquid gas transition from 
the vacuum to a baryonic crystal, where all lattice sites are occupied by baryons. Since the lattice spacing at strong coupling is maximally coarse, the degrees of freedom are on a hypercubic crystal and
saturation is due to the Pauli principle.
For staggered fermions, there is also spontaneous chiral symmetry breaking and its restoration at some critical temperature $aT_c$. In contrast to Wilson fermions, there is a remnant 
chiral symmetry $U_{55}(1)\subset SU_L(\Nf)\times SU_R(N_f)$ that is not broken by the finite lattice spacing.
The ultimate goal is to study the QCD phase diagram and the nuclear transition away from the strong coupling limit.
A first step into that direction, the $O(\beta)$ corrections to the strong coupling phase diagram, has already been undertaken \cite{Unger2014}.

A dimer/flux representation is possible for both lattice actions, but they differ qualitatively.
For staggered fermions: a partition function in the monomer-dimer representation is valid for any quark mass;
there is an exact chiral symmetry, hence it is adequate to study chiral properties (also, simulations in the chiral limit are cheap); however, staggered fermions are spinless in the strong coupling limit.
Contrast this with Wilson fermions: the flux representation involves spin, but since backtracking of fermions is not allowed, $(\Id-\gamma_\mu)(\Id+\gamma_\mu)=0$,
it poses a complicated combinatorial problem and expansion in spatial hadronic hoppings is required.
Both discretizations have very different lattice artifacts. The main motivation for this analysis is the question whether they share a
``physical'' content at strong coupling or at $\Ord{\beta}$ which could be isolated from lattice artifacts. The combinatorial perspective may help to shed light on this question.

\section{Gauge Integrals, Invariants and Combinatorics}

Combinatorics can give additional insight into lattice QCD, when formulated in a 
dual, color singlet representation based on integer variables. The combinatorial paradigm I want to utilize is the question of how many ways there are to put $n$  balls into $k$ boxes.
Many combinatorial problems reduce to this question, and the answer will depend on the permutation symmetries in the problem, i.e.~whether balls or boxes are distinguishable or not, and which restrictions on the 
placements are made (e.g.~the 12 canonical answers known as ``twelvefold way'').
Combinatorial formulae amount to integer sequences, which are listed in the On-Line Encyclopedia of Integer Sequences \cite{OEIS}. In the following I quote the A numbers from OEIS for further explanations and proofs.

First consider the SU$(\Nc)$ one-link integral \cite{Creutz1978,Eriksson1981} 
which can be evaluated both for staggered and Wilson fermions:
\begin{align}
z(x,\mu)&=\int dU_\mu(x) e^{\tr_c \left[U_\mu(x)M^\dagger + U_\mu(x)^\dagger M\right]},\\
(M^{\rm stagg.})_{ij}&=\chi^{f}_i(x)\bchi^{f}_j(x+\hmu),\qquad 
(M^{\rm Wilson})_{ij}&=\psi^{\beta,f}_{i}(x)(\Id-\gamma_\mu)_{\alpha\beta}\bpsi^{\alpha,f}_{j}(x+\hmu),
\end{align}
with $i,j\in\{1,\ldots \Nc\}$, $f\in\{1,\ldots \Nf\}$ and $\alpha$, $\beta$ Dirac indices.
In both cases, the link integrals are gauge invariants, which can be expressed by linear combinations of traces and determinants.
\begin{align}
z(x,\mu)=\sum_{k_1,\ldots k_{\Nc+1}}\alpha_{k_1\ldots k_{\Nc+1}}\det_c\left[M\right]^{k_1}\det_c\left[M^\dagger\right]^{k_2}\tr_c\left[MM^\dagger\right]^{k_3}\ldots \tr_c\left[(MM^\dagger)^{\Nc-1}\right]^{k_{\Nc+1}}
\end{align}
The prefactors $\alpha_{k_1\ldots k_{\Nc+1}}$ can be determined via Grassmann identities, e.g.~for $\Nf=1$ staggered fermions 
(with $y=x+\hmu$ and $B(x)=\frac{1}{\Nc!}\epsilon_{i_1\ldots i_{\Nc}}\chi_{i_1}(x)\ldots \chi_{i_{\Nc}}(x)$):
\begin{align}
e^{\bchi_y\chi_y}&=\int d\chi_xd\bchi_x\int dU e^{\bchi_x \chi_x+\bchi_xU \chi_y-\bchi_yU^{\dagger} \chi_x}=\sum_{l=0}^{\Nc} \alpha_k \frac{\Nc!}{(\Nc-k)!}(\bchi_x \chi_x\bchi_y \chi_y)^k\\
\Rightarrow \qquad z(x,y)&=\sum_{k=0}^{\Nc}\frac{(\Nc-k)!}{\Nc!k!}(M_x M_{y})^k +\bar{B}(x)B(y)+ (-1)^{\Nc} \bar{B}(y)B(x). 
\end{align}

\noindent \begin{minipage}{0.65\textwidth}
The prefactors can also be determined via combinatorics (labeled balls into labeled boxes, see Fig.~\ref{ballsboxes}). This strategy can be generalized to also apply to $\Nf>1$ 
and for Wilson fermions, where meson hoppings $(M_x M_{x+\hmu})$ and baryon hoppings $\bar{B}(x)B(x+\hmu)$ carry flavor and spin. 
The corresponding integrals have been determined for $\Nc\leq 3$ in \cite{Eriksson1981} and are in agreement with the combinatorial 
determination.
\end{minipage}
 \qquad
\begin{minipage}{0.3\textwidth}
\vspace{-6mm}
\hspace{-6mm}
\includegraphics[width=1.1\textwidth]{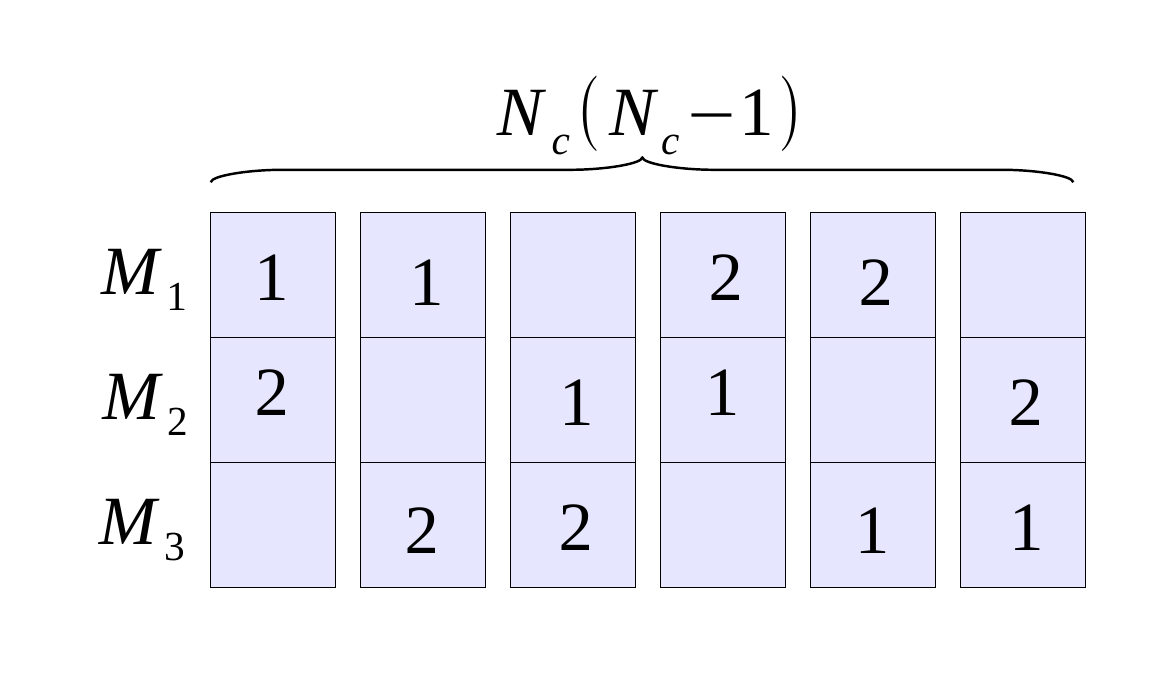}
\captionof{figure}{
\label{ballsboxes}
$k=2$ balls ($\bar{\chi}\chi$) into $\Nc=3$ boxes (mesons).\\[-2mm]}

\end{minipage}\\

\begin{table}
\begin{minipage}{0.6\textwidth}
\begin{tabular}{cccccc}
 \hline 
           &  $1$ {\tiny{(mesonic)}} & $L^{3}$ & $L^{6}$ & $L^{9}$ & $L^{12}$ \\
 \hline 
\hspace{4mm}$1$ {\tiny{(baryonic)}}\hspace{-4mm} &   1 & 1 & 5 & 42 & 462 \\
$(LL^*)$   &   1 & 3 & 21 & 210 & 2574 \\
$(LL^*)^2$ &   2 & 11 & 98 & 1122 & 15015 \\
$(LL^*)^3$ &   6 & 74 & 498 & 6336 & 91091 \\
$(LL^*)^4$ &  23 & 225 & 2709 & 37466 & 571428 \\
 \hline 
\end{tabular}
\end{minipage}
\begin{minipage}{0.4\textwidth}
\includegraphics[width=\textwidth]{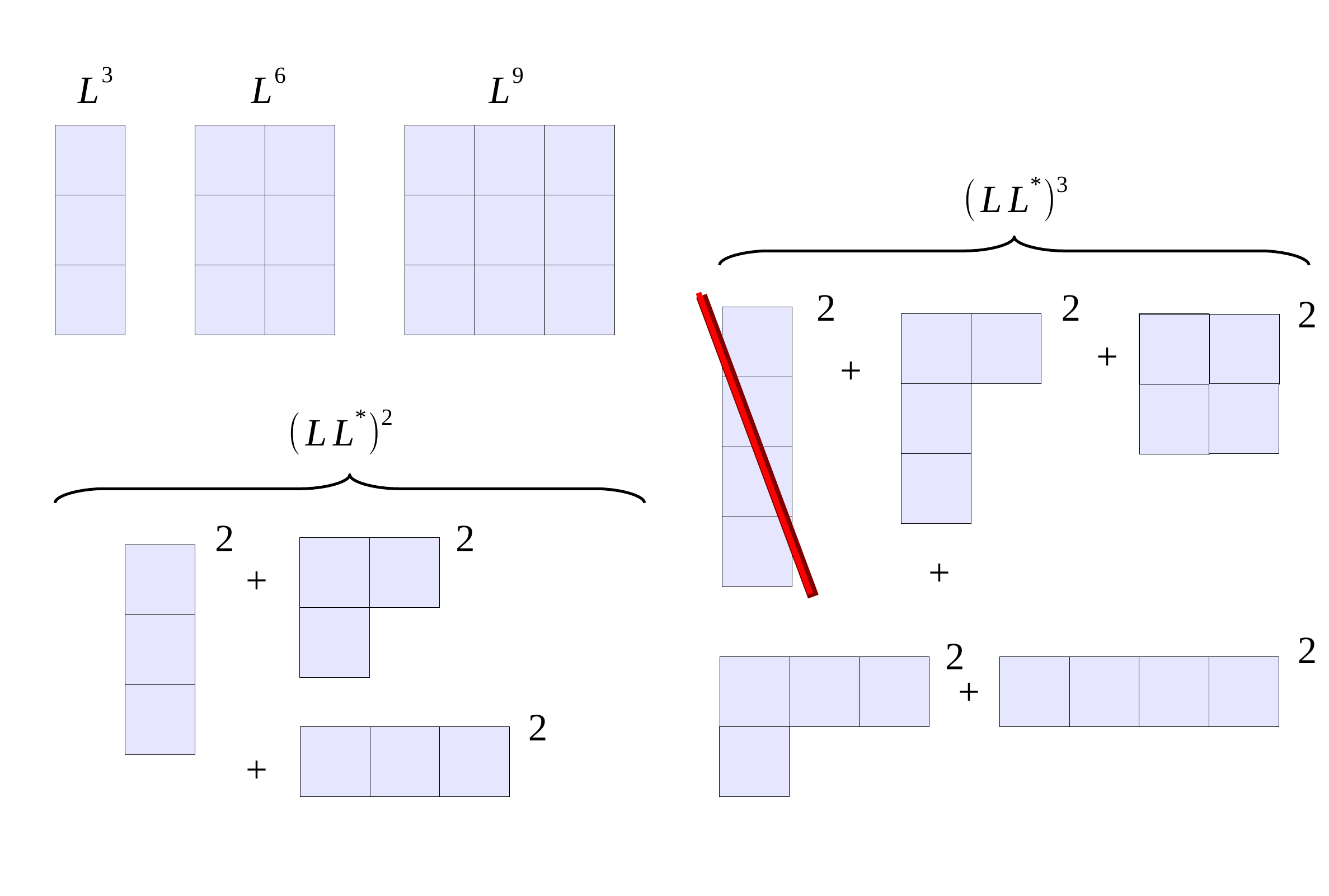}
\end{minipage}
\captionof{table}{
\label{Table}
List of SU(3) gauge integrals $\langle(LL^*)^n(L^3)^m\rangle$, 
which enumerate the number of restricted permutations patterns, which is the number representations of bounded height, see Eq.~(2.8).
}
\end{table}

Another type of gauge integrals are those over the trace of closed loops of gauge links: let 
$P=\prod\limits_{(x,\mu)\in\mathcal{C}} U_\mu(x)={\rm diag}\left(e^{i\phi_1},\ldots e^{i\phi_{\Nc-1}},e^{i\phi_{\Nc}}\right)$ with 
$\phi_{\Nc}=-\sum\limits_{k=1}^{\Nc-1}\phi_k$
be any closed loop of gauge links along contour $\mathcal{C}$ (e.g. Polyakov loop, Wilson loop), then, for SU($\Nc$) there are $\Nc-1$ gauge invariants, such as $L=\tr_c [P]$ and $L^*$.
Only the mesonic $\mathcal{M}=LL^*$, baryonic $\mathcal{B}=L^{\Nc}$, and mixed operators $\mathcal{O}=\mathcal{M}^n\mathcal{B}^m$ for $n,m\in \mathbb{N}$) have non-vanishing expectation values:
\begin{align}
\expval{\mathcal{O}(L,L^*\ldots)}=\frac{1}{(2\pi)^{\Nc-1}}\int d\phi_1 \ldots d\phi_{\Nc-1}\, V(L,L*,\ldots) \mathcal{O}(L,L^*,\ldots),
\end{align}
where  $V(L,L^*\ldots)$ is obtained from the invariant Haar measure:
$d\mu(\boldsymbol{\phi})=\prod\limits_{i>j}|e^{i\phi_i}-e^{i\phi_j}|^2\prod\limits_i d\phi_i$.
The result for SU(2), where $L^*=L$, gives rise to the so-called Catalan numbers (A000108), which play a prominent role in combinatorics, e.g. 
as the number of 123-avoiding permutation patterns:
\begin{align}
\expval{L^{2n}}=\frac{1}{2\pi}\int_0^{2\pi} d\phi 2 \sin^2 \phi (2\cos \phi)^{2n}=\frac{1}{2\pi}\int_{-2}^2 dL \sqrt{4-L^2}L^{2n}=C_n\equiv\frac{1}{n+1}{2n \choose n} 
\end{align}
For SU(3), where $\tr[P^2]=L^2-2{L^*}^2$, the various results listed in Tab.~\ref{Table} can be mapped on representations of the permutation group. 
The invariants of higher moments in $L$, $L^*$, which are needed to express the static limit of Wilson fermions (see below), can be characterized as restricted permutation patterns, 
which correspond to dimensions of standard young tableaux of bounded height.
\begin{align}
\label{bound}
m_{\Nc}(n)=\sum\limits_{h(\lambda_n)\leq \Nc} d_{\lambda_n}^2,\qquad   b_{\Nc}(n)= d_{n\times\Nc},\qquad mix_{\Nc}(n_m,n_b)=
\sum\limits_{h(\lambda_{n_m})\leq \Nc} 
d_{n_b\times \Nc,\lambda_{n_m}} d_{\lambda_{n_m}}.
\end{align}
To compute these invariants in the general case of SU($\Nc$) or U($\Nc$) and $\Nf>1$, one needs to evaluate $\tr[P^n]$, where $n=1,\ldots, \Nc\Nf$. These can be obtained via generalized Lucas polynomials.
The Lucas $n$-step numbers are $F_k^{(n)}=\sum\limits_{i=1}^{n}F_{k-i}^{(n)}$ (which is Fibonacci-like for $n=2$).
Related to SU(3) are the 3-step Lucas numbers
$F_{k}^{(3)}=F_{k-1}^{(3)}+F_{k-2}^{(3)}+F_{k-3}^{(3)}$
with seeds $F_{0}^{(3)}=3$, $F_{1}^{(3)}=1$, $F_{2}^{(3)}=3$, 
from which the following 3-step polynomials $F$ in the variables $x$, $y$, $z$ are obtained :
\begin{align}
F^{(3)}_{n}(x,y,z)=
  \tr \left[\left(
  \begin{array}{ccc}
   x & y & z\\
   1 & 0 & 0\\
   0 & 1 & 0\\
  \end{array}
  \right)^n
 \right],\quad 
 \tilde{F}^{(3)}_{n}(x,y,z)=
  \tr \left[\left(
  \begin{array}{ccc}
   x & y & z\\
   -1 & 0 & 0\\
   0 & -1 & 0\\
  \end{array}
  \right)^n
 \right].
\end{align}
It turns out that the signed verison $\tilde{F}_n$ is directly related to $\tr[P^n]=\tilde{F}_n(P)$  by identifiying $x\equiv L=\tr[P]$, $y\equiv L^*=\tr[P^\dagger]$, $z\equiv D=\det[P]$ (=1 for SU($\Nc$)). The first orders are
 \begin{align*}
 \tr[P^0]&=3, &
 \tr[P^1]&=L, \\ 
 \tr[P^2]&=L^2-2L^*, &
 \tr[P^3]&=L^3-3L L^* +3D,\\ 
 \tr[P^4]&=L^4-4L^2L^*+2L^{*2}+4LD, &
 \tr[P^5]&=L^5-5L^3L^*+5L L^{*2}+5L^2D-5L^{*}D.
 \end{align*}
The corresponding versions for arbitrary $\Nc$ is obtained by considering the signed $\Nc\times\Nc$ matrix $\tilde{F}(x_1,\ldots, x_\Nc)$.
I have used the generalized Lucas polynomials to determine the flavor dependence of the static limit.

\section{Strong Coupling Partition Functions}

The final partition functions at strong coupling are obtained after
Grassmann integration, 
which introduces site weights $w_x$. For staggered fermions:
\begin{align}
 w_x=\int\prod\limits_{c}[d\chi_{c,x}d\bchi_{c,x}]e^{2am_q\bchi_{c,x}\chi_{c,x}}(\bchi_{c,x}\chi_{c,x})^{k_x}=\frac{\Nc!}{n_x!}(2am_q)^{n_x},
\end{align}
with monomers $n_x=\Nc-k_x$, determined by the Grassmann constraint $k_x=\sum\limits_{\pm\hmu}k_{\pm\hmu}(x)$, hence \linebreak $n_x\in \{0,\ldots \Nc\}$, and no monomers at baryonic sites.
The well-known staggered partition function ($\Nf=1$) valid to all orders in the hopping parameter $\kappa=\frac{1}{2am_q}$ is
\begin{align}
Z_{SC}^{\rm stagg.}(m_q,\mu,\gamma)= \sum_{\{k_b,n_x,\ell\}}
\underbrace{\prod_{b=(x,\mu)}\frac{(\Nc-k_b)!}{\Nc!k_b!}\gamma^{2 k_b\delta_{\mu 0}}}_{\text{meson hoppings}\, M_xM_y}
\underbrace{\prod_{x}\frac{\Nc!}{n_x!}(2am_q)^{n_x}}_{
\text{chiral condensate}\, M_x}
\underbrace{\prod_\ell w(\ell,\mu)}_{\text{baryon hoppings}\, \bar{B}_xB_y},
\end{align}
with $k_b\in \{0,\ldots \Nc\}$, $n_x \in \{0,\ldots \Nc\}$, $\ell_b \in \{0,\pm 1\}$.
The weight $w(\ell,\mu)$ and sign $\sigma(\ell)=\pm 1$ for an oriented loop $\ell$ depend on loop geometry.
The anisotropy $\gamma=a/a_t$ is needed to vary the temperature continuously at $\beta=0$ \cite{Unger2012}.
\\
For Wilson fermions, Grassmann integration amounts to
spin and flavor conservation. 
The site weights (almost) cancel link weights. 
Only when spatial hoppings of color neutral states occur, the site weights are non-trivial. The partition function can generally be mapped on a vertex model. 
This has been done for the Schwinger model, which maps on a 7-vertex model for $\Nf=1$ \cite{Salmhofer1991} and on a modified 3-state 20-vertex model for $\Nf=2$ \cite{Scharnhorst1996}.
Grassmann integration for $\Nc>1$ is too complicated to do by hand but can be automatized using computer algebra.
The Wilson fermion partition function has the general structure
\begin{align}
 Z_{SC}^{\rm Wilson}(\kappa,\mu)= \sum_{\{k_b,n_x,\ell_j\}} N(\{k_b,\ell_j\}) v_i^{C_i}
\prod_x \frac{1}{(2\kappa)^{n_x}}\prod_{\ell_j} w(\ell_j,\mu).
\end{align}
$C_i$ counts how often vertices of type $i$ occur and $N(\{k_b,\ell_j\})$ counts multiplicities of loops. There are various baryonic loops $\ell_j$ (depending on spin and flavor).
The Grassmann constraint allows mesonic and baryonic world lines to intersect even for $\Nf=1$. The vertex weights $v_i$ still need to be determined in general via Grassmann integration.

\begin{figure}
\vspace{-5mm}
\begin{minipage}{0.12\textwidth}
\begin{center}
\includegraphics[width=\textwidth]{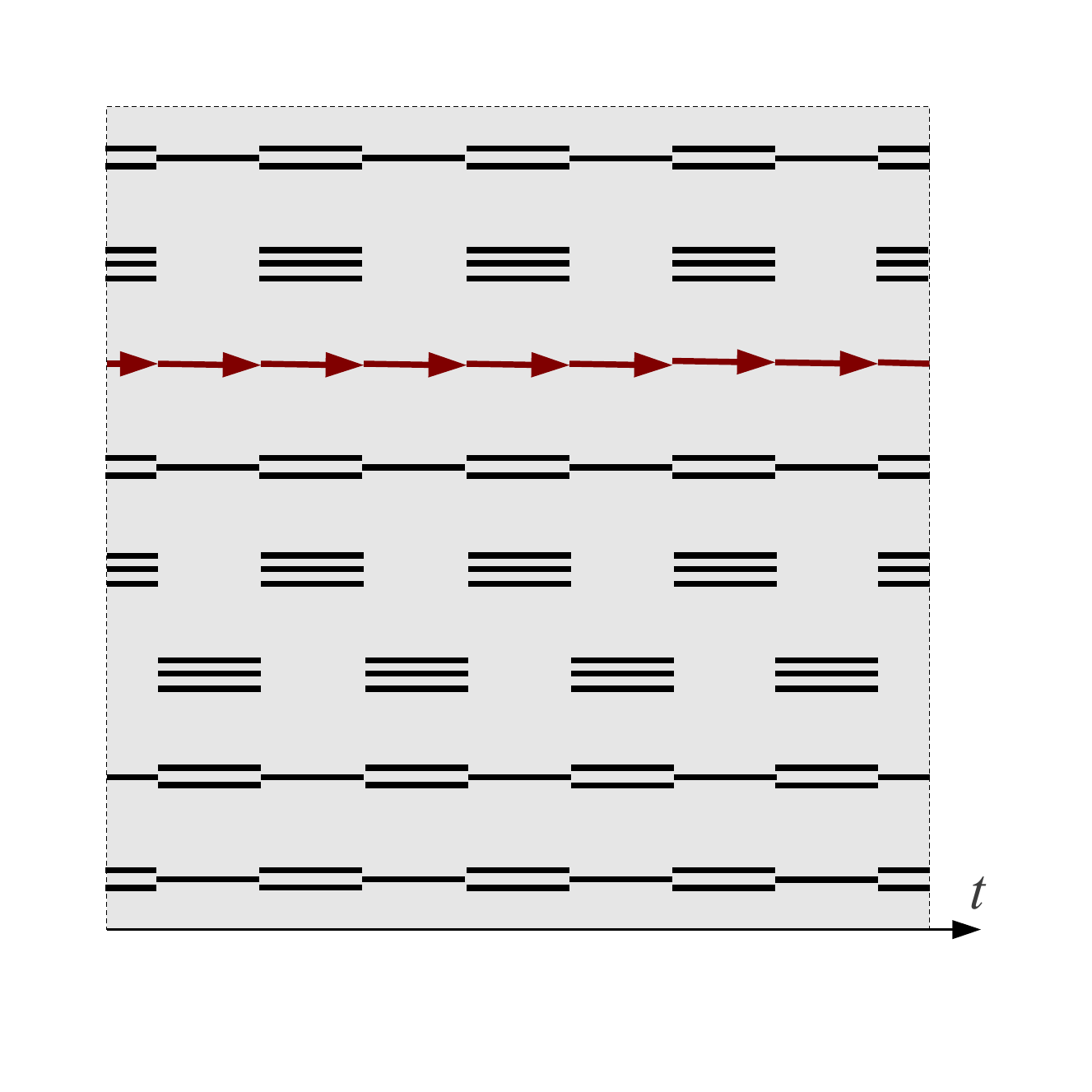}\\[-4mm]
{\scriptsize $\mu=0,T\gg T_c$ }\\[4mm]
\includegraphics[width=\textwidth]{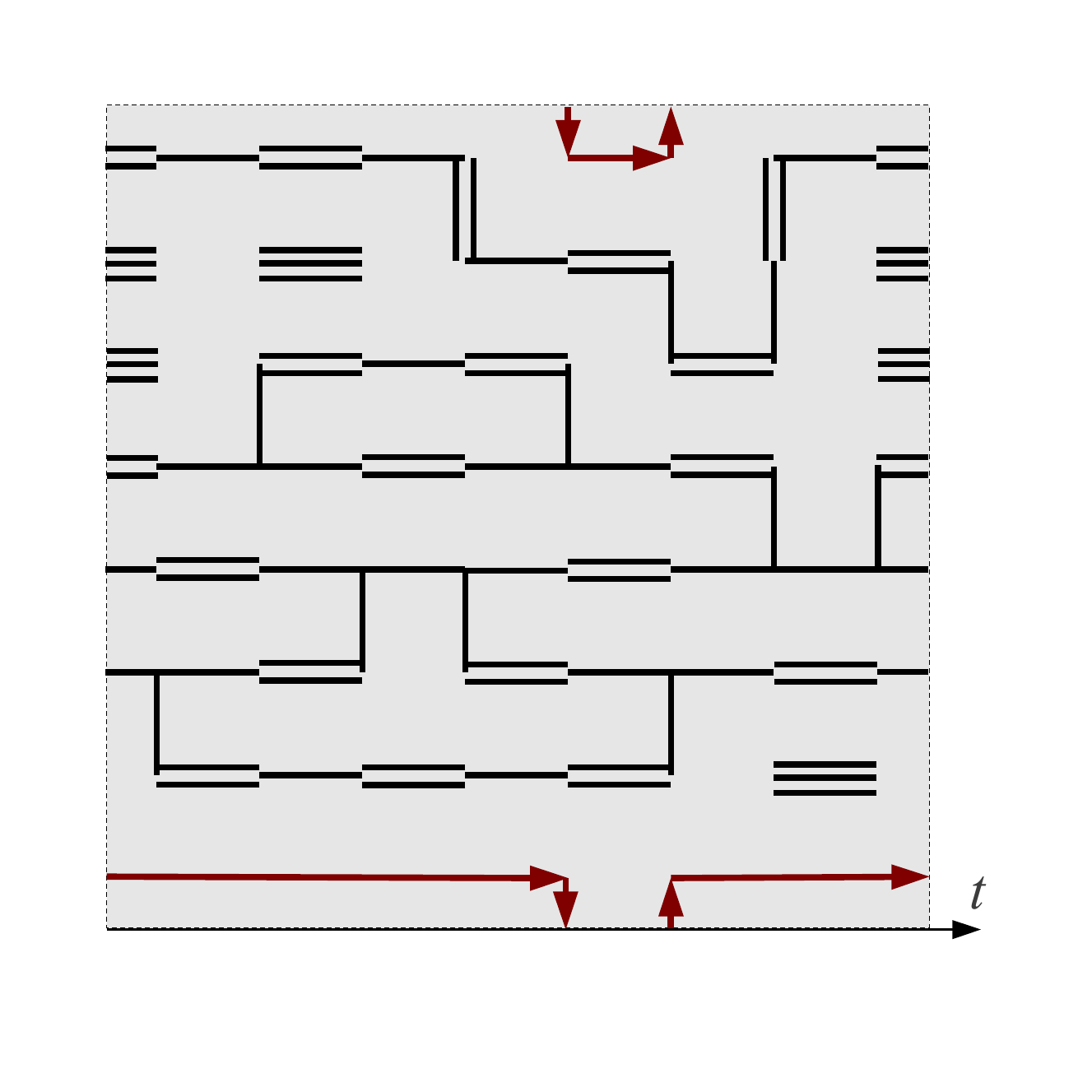}\\[-4mm]
{\scriptsize$\mu=0,T<T_c$}\\
\end{center}
\end{minipage}
\hspace{-3mm}
\begin{minipage}{0.35\textwidth}
\centerline{{\footnotesize Staggered Fermions}}
\vspace{2mm}
\includegraphics[width=\textwidth]{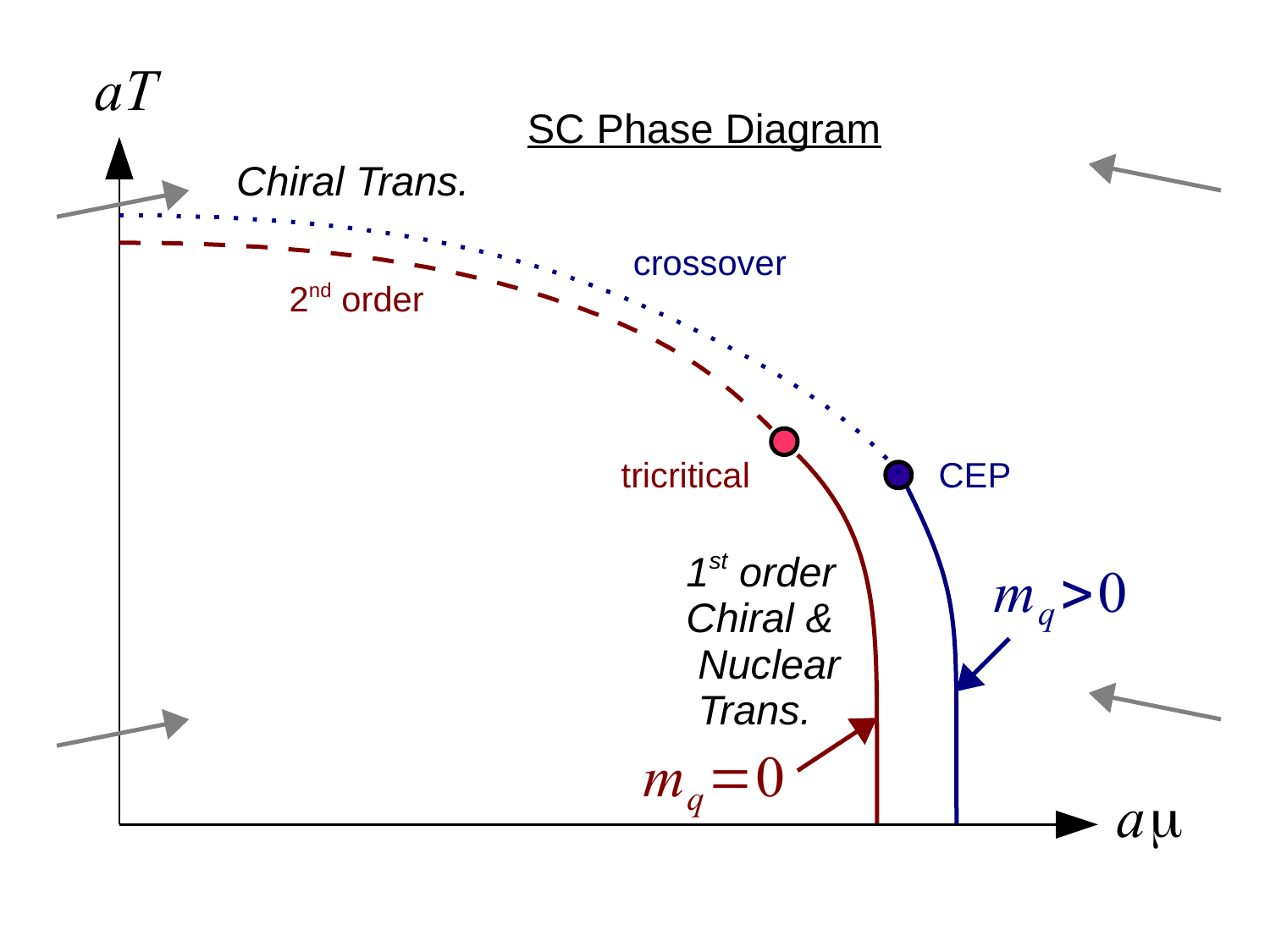}
\end{minipage}
\hspace{-3mm}
\begin{minipage}{0.12\textwidth}
\begin{center}
\includegraphics[width=\textwidth]{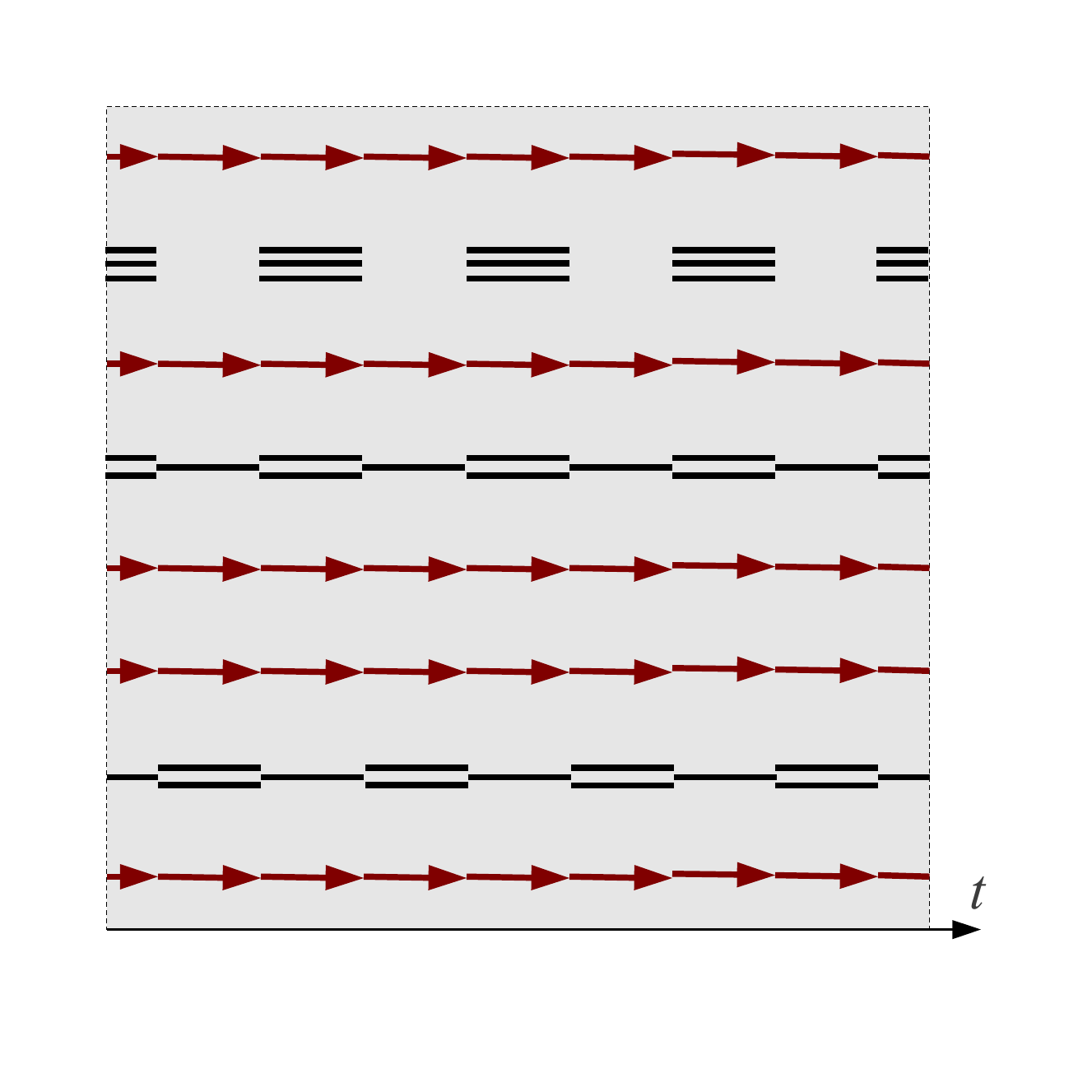}\\[-4mm]
{\scriptsize$\mu > \mu_c,\,T\gg T_c$}\\[4mm]
\includegraphics[width=\textwidth]{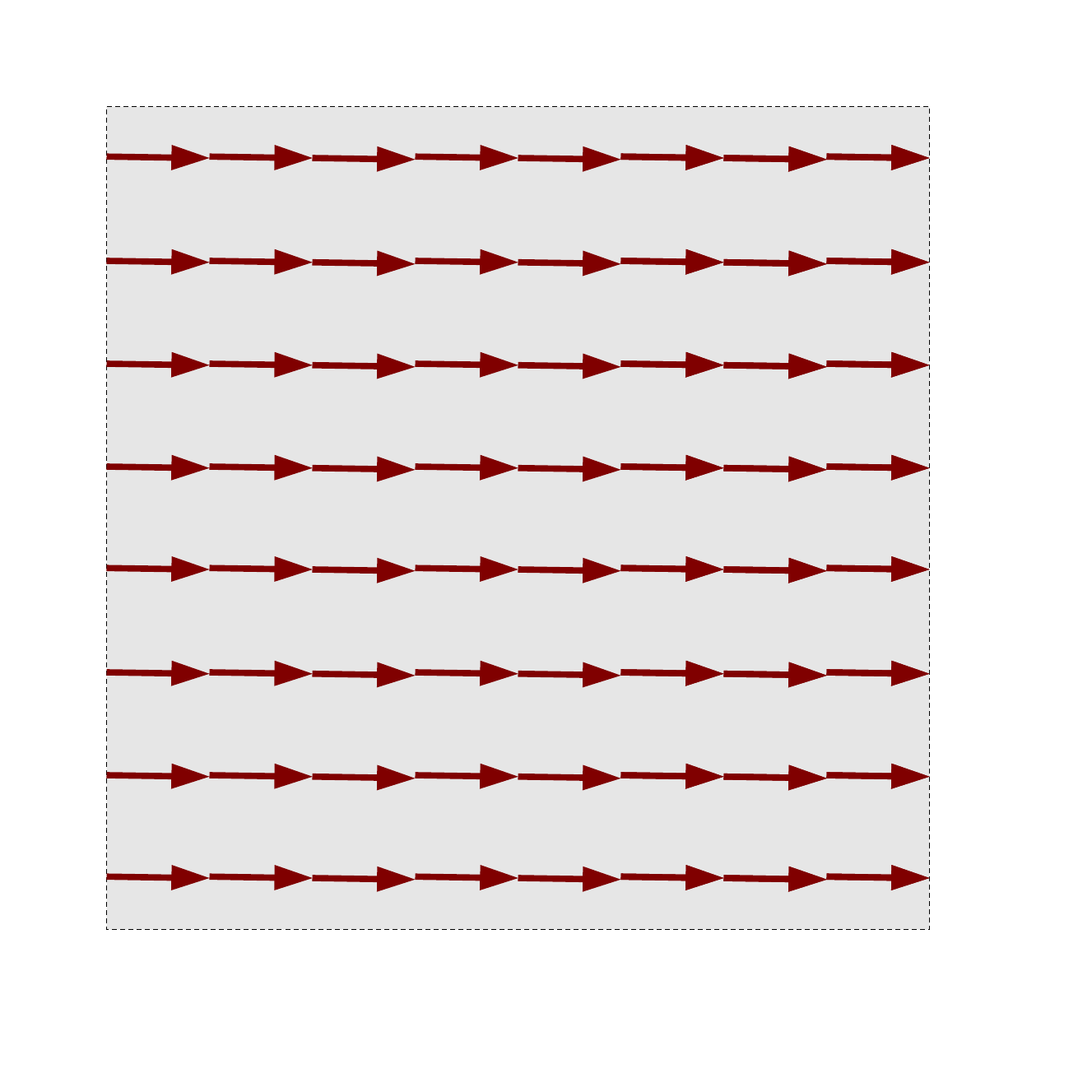}\\[-4mm]
{\scriptsize$\mu > \mu_c,\,T=0$}\\
\end{center}
\end{minipage}\;
\begin{minipage}{0.01\textwidth}
|\\[-3mm]
|\\[-3mm]
|\\[-3mm]
|\\[-3mm]
|\\[-3mm]
|\\[-3mm]
|\\[-3mm]
|\\[-3mm]
|\\[-3mm]
|\\[-3mm]
|\\[-3mm]
|\\[-3mm]
|\\[-3mm]
|\\[-3mm]
|\\[-3mm]
|\\[-3mm]
|\\[-3mm]
|\\[-3mm]
|\\[-3mm]
|\\[-3mm]
\end{minipage}
\begin{minipage}{0.4\textwidth}
\vspace{4mm}
\centerline{{\footnotesize Static Wilson Fermions}}
\vspace{3mm}
\begin{minipage}{0.49\textwidth}
\includegraphics[width=\textwidth]{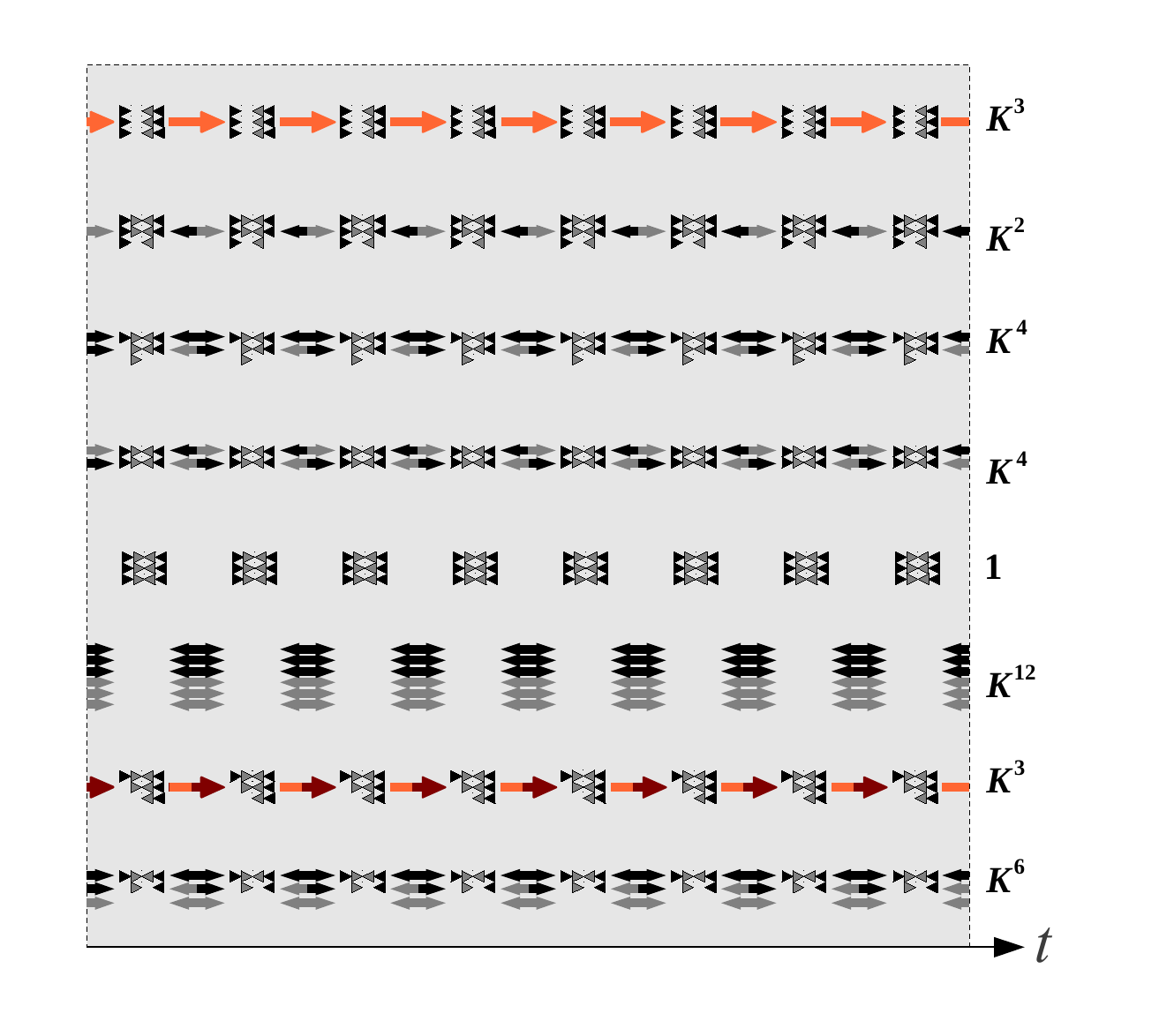}\\[-3mm]
\centerline{{\footnotesize various $\mathcal{O}(K_t)$}}\\[4mm]
\end{minipage}
\begin{minipage}{0.49\textwidth}
\includegraphics[width=\textwidth]{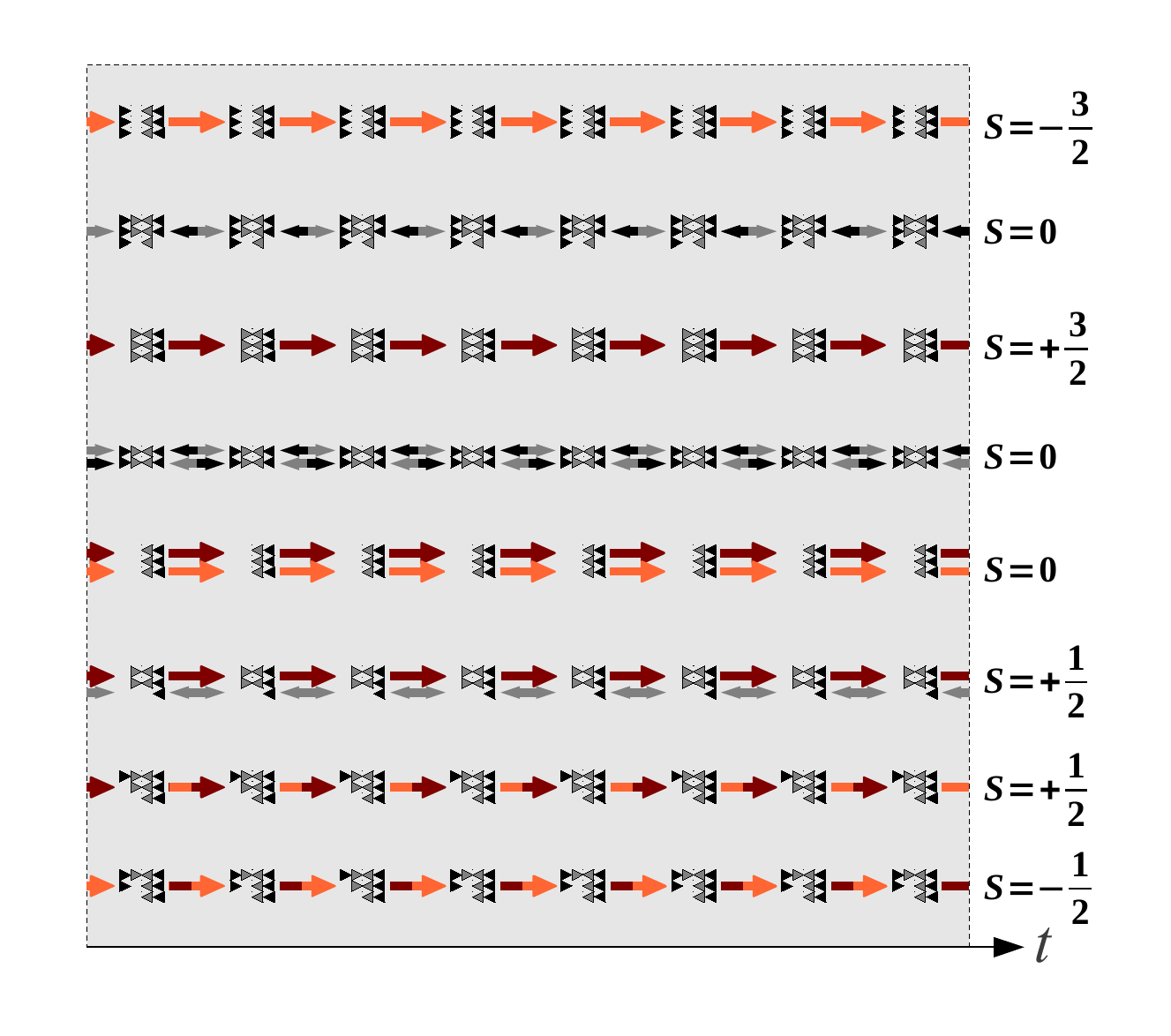}\\[-3mm]
\centerline{{\footnotesize various spins}}\\[4mm]
\end{minipage}
\end{minipage}\hspace{-3mm}
\vspace{-2mm}
\caption{
\label{Configs}
At $\beta=0$, both staggered and Wilson fermions become static at large $T$ and/or $\mu$. Left: The staggered SC phase diagram, it has been measured in [4,7].
Right: Static Wilson fermions described by monomers/dimers/fluxes, multiplicities are due to spin and $K_t$.
}
\end{figure}

In the static limit, i.e. in the absence of spatial fermion hoppings, the strong coupling partition function is $Z_{SC}^{\rm static}=\prod\limits_{\vec{x}}Z_1(\vec{x})$, where $Z_1$ is the sum over all possible hadronic quantum states
$|\psi\rangle$.
This describes SC-LQCD in the high temperature and/or high density regime, see Fig.~\ref{Configs}. 
For staggered fermions,
the chiral restoration takes place when the number of spatial dimers reaches a critical value. The nuclear and chiral transition coincide, because
$\expval{\bar{\chi}\chi}$ vanishes as a baryonic crystal forms. The number of hadronic states $|\psi\rangle=|P_u,P_d,\ldots P_\Nf,Q_{\pi^+},Q_{K^+},\ldots\rangle$ is
\begin{align}
\qquad Z_1(\mu,T)=\quad &{2\Nf \choose \Nc\Nf}_{\Nc}+ \sum\limits_{n=1}^{N_t\Nc/2} t_n (2am_q)^{2n}+2
{{\Nc+\Nf-1} \choose {\Nf-1}}
\cosh(\mu_B/T),
\end{align}
where the terms $\mathcal{O}(2am_q)$ are suppressed at high $T$ (for $\Nc=3$ in the chiral limit, the prefactors $t_n$ are related to Tribonacci numbers (A000073), a generalization of Fibonacci).
The degeneracies of the mesonic states are given by central polynomial coefficients (see A077042) which number the possibilities to put $n$ unlabeled balls into $k$ labeled boxes,
allowing at most $\Nc$ balls in each box.

For Wilson fermions, 
with $K_t=(2\kappa)^{\Nt}$ and hopping parameter $\kappa=\frac{1}{2d+2am_q}$, for $\Nf=1$:
\begin{align}
\label{WilsonStat}
Z_1(\mu,T)= \sum\limits_{k=0}^{2\Nc} T(k) K_t^{2k} +
\sum\limits_{k=0}^{\Nc}P(k)K_t^{(2k+\Nc)} 2\cosh(\mu_B/T)+ 
K_t^{2\Nc} 2\cosh(2\mu_B/T).
\end{align}
The combinatorics of the mesonic sector is given by the so-called tetrahedal numbers (A133826) \linebreak $T(k)=\sum\limits_{q=0}^{k}d_{D^{0q}}={3+\min(k,2\Nc-k) \choose 3 }$, with $D^{0q}$ the mesonic irreducible representations of SU($\Nc$),
and product numbers $P(k)=(1+k)(1+\Nc-k)$. This can be generalized for $\Nf>1$, e.g.\linebreak $T(k)={(2\Nf)^2-1+\min(k,2\Nc-k) \choose (2\Nf)^2-1 }$, and Eq.~(\ref{WilsonStat}) contains $2\Nf+1$ sums $\sim K_t^{2k+n\Nc}$.
To conclude, the quantum number degeneracies of all static states can be listed via combinatorial formulae.

There are two kinds of corrections to the static limit, which can be both addressed systematically via an expansion: 
(1) The (spatial) hopping parameter expansion in $\kappa_s$ allows to approach 
the chiral limit, with the number of spatial mesonic and baryonic hoppings being controlled by the quark mass.
In a finite volume, this expansion always terminates due to the Grassmann constraint!
(2) The expansion in $\beta$ (the inverse gauge coupling) allows to approach the continuum limit. The staggered strong coupling partition function is in fact valid for all quark masses 
(with the chiral limit being cheapest when addressed with a worm algorithm), whereas the Wilson partition function is restricted to rather large quark masses. 
In both lattice discretizations, the gauge action can be incorporated order by order, which gives rise to higher order link integrals.

The strategy to study both lattice discretizations on a par is to expand around the static limit by making use of the Hamiltonian formulation that can be derived in the continuous time limit, 
$\Nt\rightarrow \infty$ \cite{Unger2012}. 
In this limit, the partition function simplifies further as only single meson hoppings need to be considered. The static lines are the in and out states of the transfer matrix:
\begin{align}
Z&=\Tr[e^{\beta \mathcal{H}}],& \mathcal{H}&=
\frac{1}{2}\sum_{\langle x,y \rangle}
\sum_{Q_i}J^+_{Q_i(x)} J^-_{Q_i(y)},&
J^{-}_{Q_i}&=(J^{+}_{Q_i})^\dagger,
\end{align}
where the generalized quantum numbers $Q_i$ (spin, parity, flavor) are globally conserved, and spatial dimers represented by
$J^+_{Q_i(x)} J^-_{Q_i(y)}$ raise quantum number $Q_i$ at site $x$ and lower them at a neighboring site $y$ (see \cite{Unger2012} for the case of 
$\Nf=1,2$ for staggered fermions). For both staggered and Wilson fermions, the matrices $J^{\pm}_{Q_i}$ contain vertex weights. They are the crucial input to sample 
the corresponding partition function with a quantum Monte Carlo algorithm to all orders in $\kappa_s$, e.g.~via a continuous time Worm algorithm.
However, whether also for Wilson fermions in four dimensions all vertex weights are positive to evade the sign problem is still an open question, although due to the continuous time limit, only a small
set of vertices need to be considered. For the simulation of the Schwinger model with Wilson fermions at strong coupling with the Worm algorithm see \cite{Wenger2008}.

\begin{figure}[t!]
\vspace{-5mm}
\hspace{-7mm}
\includegraphics[width=0.58\textwidth]{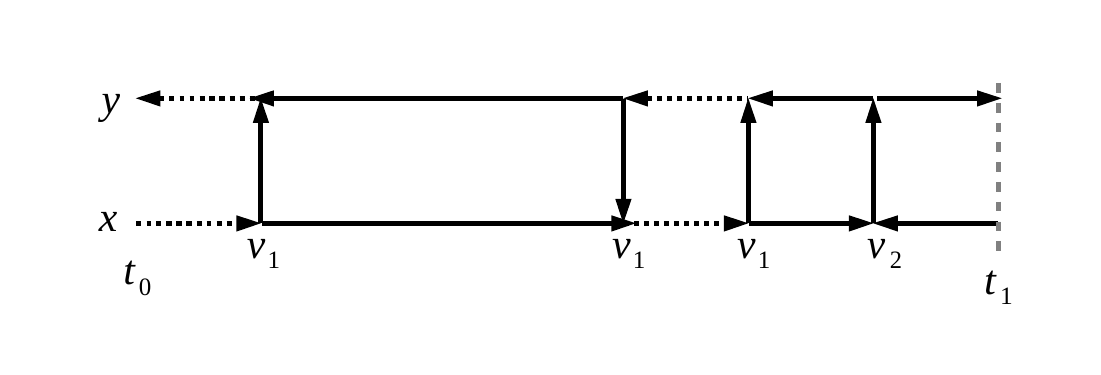}\qquad 
\vspace{-28mm}\\
\vphantom{.}\hspace{0.53\textwidth}
\includegraphics[width=0.5\textwidth]{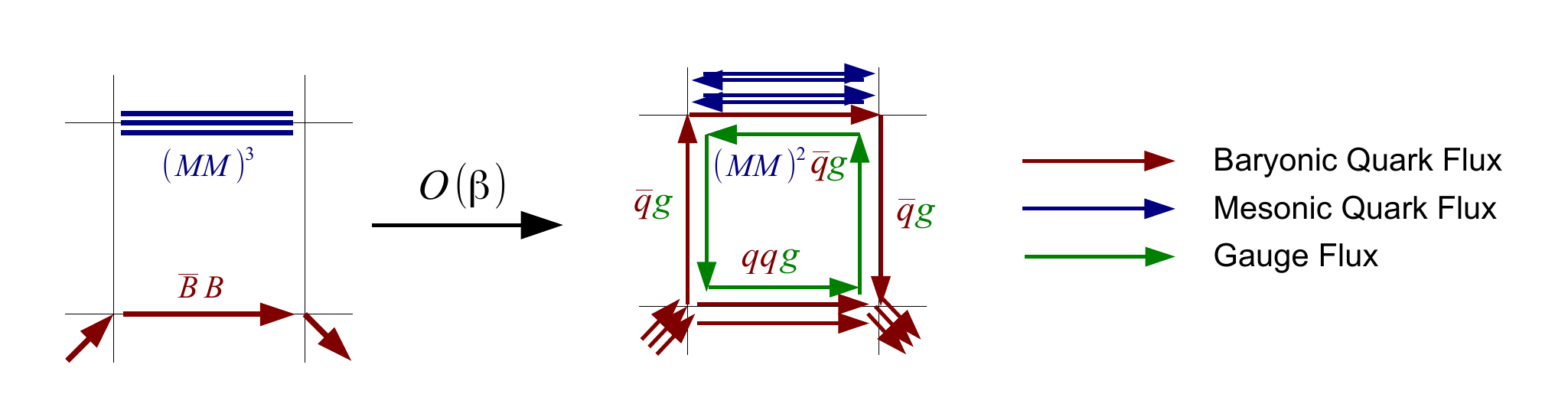}
\vspace{-5mm}
\caption{\emph{Left:} Systematic expansion in the spatial hoppings in the Hamiltonian formulation, where hadrons are emitted/absorbed at events in continuous time, depending on vertex weights $v_i$. 
\emph{Right:} Example of a gauge correction to the SC-limit, hadrons become extended objects.}
\end{figure}
Also the gauge corrections could be included in this Hamiltonian formulation.
So far, the gauge corrections have been studied for finite $\Nt=4$ and for $\Nf=1$ staggered fermions.
In a collaboration with J.~Langelage, P. de Forcrand and  O.~Philipsen, we have determined the phase diagram of SC-LQCD at $\mathcal{O}(\beta)$ \cite{Unger2014},
where the gauge action is linearized
and a new set of one-link integrals (those along an excited plaquette) have to be evaluated and new invariants with a combinatorial interpretation arise.
We find that the second order phase boundary in the $\mu$-$T$ plane is shifted to lower temperatures with increasing $\beta$, but that the tricritical point and the first order transition is invariant at $\mathcal{O}(\beta)$.
In contrast, the critical endpoint of the first order nuclear transition, which coincides with the chiral transition at $\beta=0$, moves down along the chiral first order line.
  
\emph{Acknowledgement} - I would like to thank Philippe de Forcrand and Owe Philipsen for numerous discussions. 
This works was supported by the Helmholtz International Center for FAIR
within the LOEWE program launched by the State
of Hesse.

\end{document}